\documentclass[
 prl,
 amsmath,amssymb,
 twocolumn,
superscriptaddress,
]{revtex4-2}

\usepackage{bm,graphicx,dcolumn,amsmath,amssymb,color,mathrsfs,titlesec,enumerate,float,comment,todonotes,soul,siunitx}

\usepackage{hyperref}
\definecolor{linkColor}{rgb}{1,0,0}
\hypersetup{pdfborder={0 0 0},colorlinks=true,urlcolor=linkColor,citecolor=linkColor}

\setcitestyle{authoryear,round}
\renewcommand{\thefigure}{\arabic{figure}}

\usepackage{xargs} 

\newcommand{\ii}{{\mathnormal{i}\,}}

\emergencystretch=\maxdimen
\begin{document}

\title[MBED]{Multibeam Electron Diffraction}

\author{Xuhao Hong}
\affiliation{School of Physics, Nanjing University, Nanjing, PR China 210093}

\author{Steven E Zeltmann}
\affiliation{Department of Materials Science and Engineering, University of California, Berkeley, USA, 94720}

\author{Benjamin H Savitzky}
\affiliation{National Center for Electron Microscopy, Molecular Foundry, Lawrence Berkeley National Laboratory, 1 Cyclotron Road, Berkeley, CA, USA, 94720}

\author{Luis Rangel DaCosta}
\affiliation{National Center for Electron Microscopy, Molecular Foundry, Lawrence Berkeley National Laboratory, 1 Cyclotron Road, Berkeley, CA, USA, 94720}

\author{Alexander M\"{u}eller}
\affiliation{National Center for Electron Microscopy, Molecular Foundry, Lawrence Berkeley National Laboratory, 1 Cyclotron Road, Berkeley, CA, USA, 94720}

\author{Andrew M Minor}
\affiliation{National Center for Electron Microscopy, Molecular Foundry, Lawrence Berkeley National Laboratory, 1 Cyclotron Road, Berkeley, CA, USA, 94720}
\affiliation{Department of Materials Science and Engineering, University of California, Berkeley, USA, 94720}

\author{Karen Bustillo}
\affiliation{National Center for Electron Microscopy, Molecular Foundry, Lawrence Berkeley National Laboratory, 1 Cyclotron Road, Berkeley, CA, USA, 94720}

\author{Colin Ophus}
\email{cophus@gmail.com}
\affiliation{National Center for Electron Microscopy, Molecular Foundry, Lawrence Berkeley National Laboratory, 1 Cyclotron Road, Berkeley, CA, USA, 94720}

\date{\today}
\begin{abstract}

One of the primary uses for transmission electron microscopy (TEM) is to measure diffraction pattern images in order to determine a crystal structure and orientation.  In nanobeam electron diffraction (NBED) we scan a moderately converged electron probe over the sample to acquire thousands or even millions of sequential diffraction images, a technique that is especially appropriate for polycrystalline samples. However, due to the large Ewald sphere of TEM, excitation of Bragg peaks can be extremely sensitive to sample tilt, varying strongly for even a few degrees of sample tilt for crystalline samples. In this paper, we present multibeam electron diffraction (MBED), where multiple probe forming apertures are used to create mutiple STEM probes, all of which interact with the sample simultaneously. We detail designs for MBED experiments, and a method for using a focused ion beam (FIB) to produce MBED apertures. We show the efficacy of the MBED technique for crystalline orientation mapping using both simulations and proof-of-principle experiments. We also show how the angular information in MBED can be used to perform 3D tomographic reconstruction of samples without needing to tilt or scan the sample multiple times. Finally, we also discuss future opportunities for the MBED method.

% Thus NBED measurements can easily miss a lot of the potential diffraction signal. 

%One method to overcome this limitation is precession electron diffraction, where the STEM beam is rotated over a cone of 

\end{abstract}
\pacs{}
\keywords{Diffraction, Multibeam Electron Diffraction (MBED), Four Dimensional-Scanning Transmission Electron Microscopy (4D-STEM), Amplitude Aperture}
\maketitle

\section{Introduction}
\label{section:intro}

In transmission electron microscopy (TEM), a coherent beam of high-energy electrons is used to illuminate a sample. One of the most commonly used operating TEM modes is diffraction, where the electron wave is imaged in the far-field after scattering. This experiment provides a great deal of information about the sample structure, especially for crystalline samples \citep{degraef2003introduction, zuo2017advanced, zhang2020short}. Electron diffraction of nanoscale regions is a powerful characterization technique for several reasons: the strong interaction of free electrons with matter \citep{shull1948x}, the ability to tune the sample-interaction volumes from micrometer to sub-nanometer length scales \citep{bendersky2001electron}, the ease of collecting information at high scattering angles including higher order Laue zones \citep{bird1989theory}, and the relative ease of the interpretation of electron diffraction patterns due to their small wavelengths, at least for thin specimens \citep{vainshtein2013structure}.

\begin{figure}[htbp]
    \centering
    \includegraphics[width=3.4in]{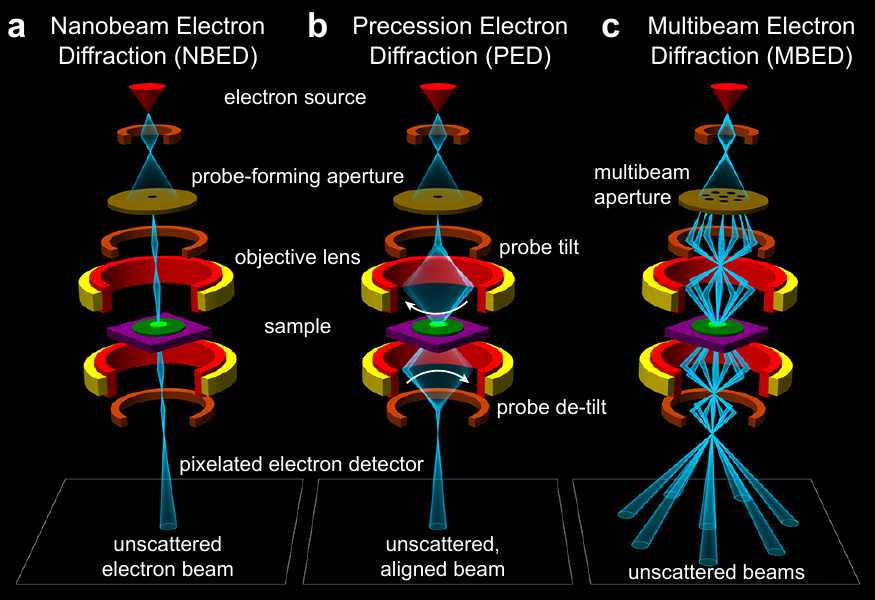}
    \caption{{\bf Experimental layout of electron diffraction experiments.} (a) Nanobeam electron diffraction, (b) precession electron diffraction, and (c) multibeam electron diffraction.}
    \label{Fig:Schematic}
\end{figure}

The widespread adoption of modern electron detector technology enables thousands or even millions of scattered electron images to be collected in a single experiment. When combined with scanning transmission electron microscopy (STEM), we can record images of the sample in reciprocal space over a two-dimensional grid of probe positions, generating a four dimensional output often referred to as a 4D-STEM experiment \citep{ophus2014_4DSTEM}. One subset of 4D-STEM experiments is nanobeam electron diffraction (NBED), where nanometer-sized electron probes are recorded in diffraction space in order to characterize the sample's structure \citep{smeaton2020mechanistic}, degree of short- or medium-range order \citep{martin2020detection}, local crystal orientation \citep{panova2019diffraction}, strain deformation fields \citep{ozdol2015strain, hughes2019relationship}, or local electromagnetic fields \citep{bruas2020improved}. A review of the various 4D-STEM measurement modes is given by \cite{ophus2019_4DSTEM_review}. The experimental layout of a conventional NBED experiment is shown in Fig.~\ref{Fig:Schematic}(a). 

A major challenge in the interpretation of NBED data is the high sensitivity of the recorded CBED patterns to slight mistilt of the sample from the zone axis \citep{chen2020challenges} and to dynamical diffraction effects \citep{donatelli2020inversion, mawson2020suppressing} in all but the thinnest samples. The presence of multiple crystallites that are not all oriented for strong diffraction also complicates analysis, as many regions of the sample may have an ambiguous interpretation or almost no diffraction signal at all. Traditionally, these challenges could be overcome though precession electron diffraction (PED), shown in Fig.~\ref{Fig:Schematic}(b), where the incident beam is precessed about the optical axis by deflector coils as the diffraction pattern is recorded \citep{rouviere2013improved}. This has the effect of incoherently averaging the diffraction patterns from each angle in the precession cone, suppressing dynamical effects and potentially exciting diffraction conditions in off-axis crystals. While this technique has been shown to be highly beneficial in many NBED measurements, particularly strain mapping \citep{mahr2015theoretical}, it requires painstaking alignment and often necessitates additional hardware in order to combine precession with STEM.

In this work, we present an alternative approach using a custom-fabricated amplitude grating in the microscope condenser optics to produce several tilted beams which all simultaneously illuminate the sample and are recorded in parallel as shown in Fig.~\ref{Fig:Schematic}(c), which we call multibeam electron diffraction (MBED). Using this unique condenser configuration we produce seven off-axis beams, tilted at up to 60~mrad (approx. 3.5$^\circ$) with respect to the optical axis, plus one on-axis beam, which are focused to the same location on the sample and record diffraction from each simultaneously. From this data we compute orientation maps of polycrystalline samples and find substantially improved coverage of reciprocal space as compared to a typical NBED experiment. Since this technique also retains diffraction information from each incident beam direction independently, we further utilize the data to compute tomographic reconstructions of the 3D structure of the sample. Finally, we discuss possible extensions to the MBED technique.

\section{Background and Theory}

\subsection{Multibeam Electron Sources}

In this study, we have generated multiple electron probes using a set of spatially-separated, fully-open apertures situated in in the probe-forming condenser plane. There are alternative methods for creating multiple STEM beams such as the two flavors of STEM holography: using diffraction gratings \citep{harvey2018stemh, yasin2018stemh}, or using a biprism to split the beam \citep{cowley2003ultra, cowley2004off}. However, diffraction gratings are very difficult to implement with small probe-forming apertures (small $q_{\textup{max}}$) due to the very fine grating pitch required to achieve large beam anglular separation. Additionally, this technique does now allow for a spatially-separated measurement of the individual beams in the diffraction plane, as the beams will recombine in the far field. Biprism beam splitters have even more limitations for MBED applications: the biprism will block part of the beam and cause diffraction at the edges, multiple biprisms are required to form more than 2 beams, and only powers of 2 number of total beams can be generated. 

Multibeam experiments similar to the experiments described in this paper are more common in scanning electron microscopy (SEM) and electron beam lithographic patterning instruments. We have taken the same approach for MBED as multibeam SEM, where a single electron source is divided into multiple beams using a patterned aperture \citep{mohammadi2010multibeam}. These systems have demonstrated much higher SEM imaging throughput due to their parallel capabilities \citep{eberle2015multiple, kemen2015multi}. Similarly, using multiple beams for electron lithography can also improve throughput \citep{sasaki1981multibeam, yin2000electron}. Multibeam systems have also been implemented for electron-beam induced deposition \citep{van2006multibeam}.

\subsection{MBED Experiment Design}

% \subsection{MBED Experiment Design Without Aberration Correction}

In a typical STEM or electron diffraction experiment, the initial probe wavefunction $\Psi_0(\mathbf{q})$ is defined as a function of the diffraction space coordinates $\mathbf{q} = (q_x,q_y)$ as
\begin{equation}
    \Psi_0(\mathbf{q}) = 
        \textup{A}(\mathbf{q})
        \exp \left[
            - \ii \chi (\mathbf{q})
            \right],
\end{equation}
where $\textup{A}(\mathbf{q})$ is an aperture function defined by
\begin{equation}
    \textup{A}(\mathbf{q}) = 
    \begin{cases}
        1   &  \text{if } |\mathbf{q}| \leq q_{\textup{max}} \\
        0   &  \text{if } |\mathbf{q}| > q_{\textup{max}},
    \end{cases}
\end{equation}
where $q_{\textup{max}}$ is the maximum scattering vector contained in the probe, and $\chi (\mathbf{q})$ defines the wavefront phase shifts of the electron probe due to coherent wave aberrations. If we neglect all aberrations except for defocus $C_1$ and 3$^{\textup{rd}}$ order spherical aberration $C_3$, the coherent wave aberrations are \citep{kirkland2020advanced}
\begin{equation}
    \label{eq:wave_aberrations}
    \chi (\mathbf{q}) = 
    \pi \lambda |\mathbf{q}|^2 C_1 + 
    \frac{\pi}{2} \lambda^3 |\mathbf{q}|^4 C_3,
\end{equation}
where $\lambda$ is the electron wavelength. The scattering angles $\bm{\alpha}$ of an electron wave are related to the scattering vector by the expression
\begin{equation}
    \bm{\alpha} = \lambda \, \mathbf{q}.
\end{equation}

\begin{figure}[htbp]
    \centering
    \includegraphics[width=3.4in]{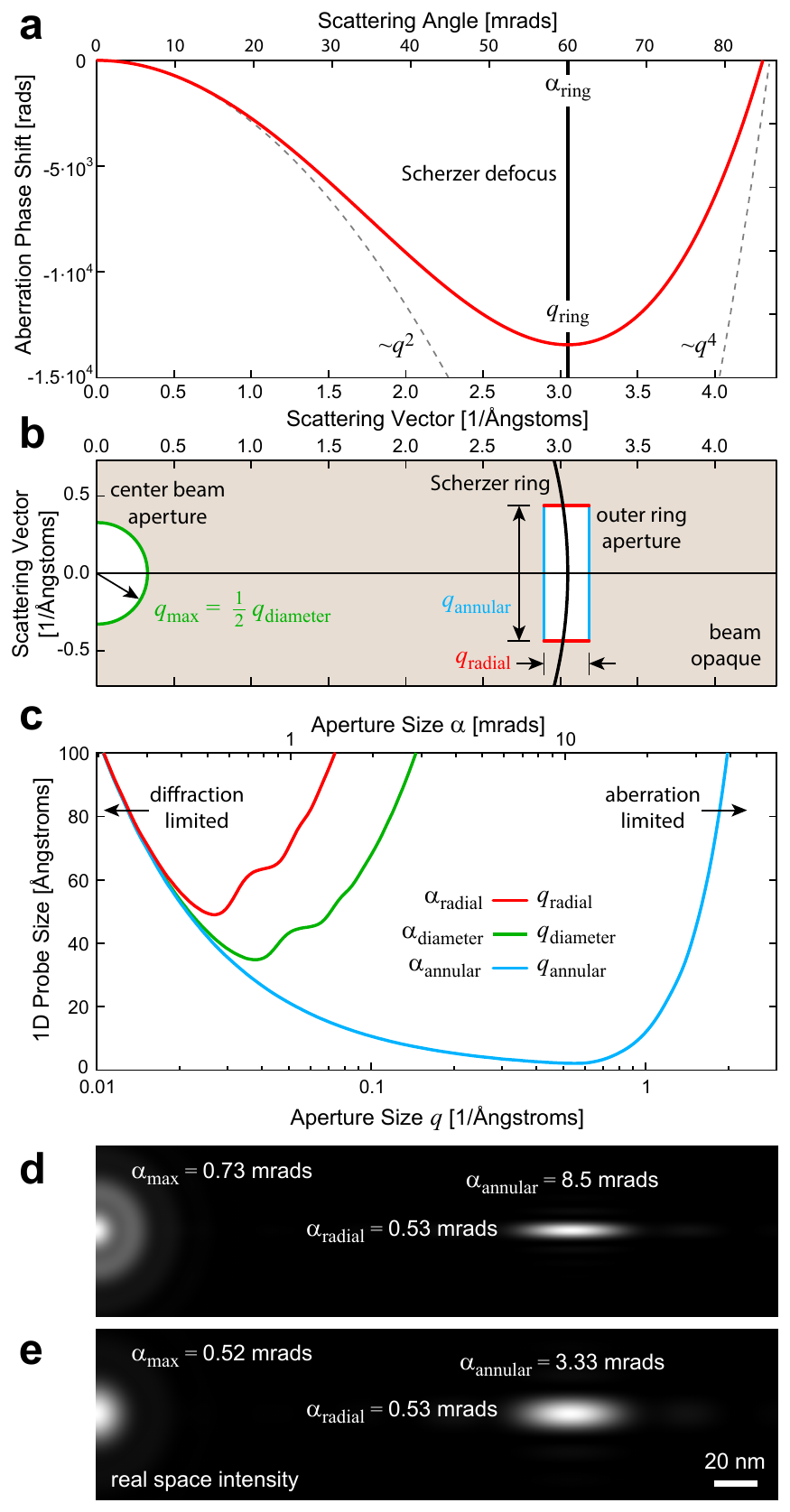}
    \caption{{\bf MBED experimental design for outer ring of probes positioned at 60 mrad, for 300 kV.} (a) Phase shifts due to defocus and 3rd order spherical aberration. (b) Aperture geometry. (c) 1D probe size as a function of aperture dimensions. (d,e) Simulated probe intensities in real space, for two geometries where outer probe intensities are matched to the center probe, including aberrations shown in (a).}
    \label{Fig:Probes}
\end{figure}

We can model a multibeam experiment by modifying the aperture function $\textup{A}(\mathbf{q})$ to contain additional non-zero regions where electrons are allowed to pass through. However, if additional electron beams are created with additional apertures at large scattering angles, these beams can be strongly affected by the coherent wave aberrations. In particular, spherical aberration causes high angle rays to be focused to a different height than ones near the optical axis. This leads to the outer beams of the MBED probe intersecting the sample plane at a different location than the central beam. To compensate for this effect of spherical aberration on the positions of the beams we operate at a ``Scherzer defocus'' condition, where a negative $C_1$ is used to balance the positive $C_3$ of a TEM without aberration correction \citep{scherzer1949theoretical}. This process is depicted schematically in Fig.~\ref{Fig:Probes}a, where we choose a defocus $C_1$ so as to yield the flattest phase error across the angle range of the outer beam. The effect of defocus on the alignment of the beams at the sample plane is shown in Fig.~\ref{Fig:apertures}c, where we label the Scherzer condition that causes all eight beams to overlap as $\Delta f = 0$.

For the initial MBED design, we used both a central probe-forming aperture and a ring of apertures at a scattering angle $\alpha_{\textup{ring}}$, which corresponds to a scattering vector $\lambda \, q_{\textup{ring}}$. The Scherzer condition requires we set $C_1$ such that the derivative of the aberration function $\chi (\mathbf{q})$ with respect to $|\mathbf{q}|$ is zero at $|\mathbf{q}| = q_{\textup{ring}}$, which is met when
\begin{equation}
    C_1 = - \lambda^2 {q_{\textup{ring}}}^2 C_3.
\end{equation}
Plugging this expression into Eq.~\ref{eq:wave_aberrations} yields
\begin{equation}
    \chi (\mathbf{q}) = 
    \pi \lambda^3 |\mathbf{q}|^2 C_3 
    \left(
        \frac{1}{2}  |\mathbf{q}|^2 - {q_{\textup{ring}}}^2
    \right).
    \label{eq:Scherzer}
\end{equation}
This expression is plotted in Fig.~\ref{Fig:Probes}a, for $\lambda = 0.0197 \, \textup{\AA}$, (300 kV), $C_3 = 1.3$ mm, and $\alpha_{\textup{ring}} =$~60 mrads.  Fig.~\ref{Fig:Probes}b shows the 2D geometry of the apertures. The center beam aperture on the optical axis has a circular shape with a radius $q_{\textup{max}}$. The outer ring apertures are positioned on the Scherzer ring. They have have a rectangular shape, with the sides parallel to the radial and annular directions. In the radial direction, the aperture size $q_{\textup{radial}}$ needs to be relatively narrow in order to form the smallest possible probe due to the steep curvature of the aberration function along this direction. By contrast, in the annular direction the aperture size $q_{\textup{annular}}$ can be made much larger due to the low curvature of the aberration phase surface along this direction.

Fig.~\ref{Fig:Probes}c shows the dependence of real-space probe size on the the 3 relevant aperture dimensions shown in Fig.~\ref{Fig:Probes}b. The probe size was estimated by first taking 1D inverse Fourier transforms in the relevant direction for a phase surface defined by Eq.~\ref{eq:Scherzer}. Next, the size was measured using the 80\% intensity criterion given by \cite{zeltmann2019patterned}. We see that there is an ideal aperture size for all three curves, which will form the smallest probe in real space. All three are diffraction-limited for small dimensions, and aberration-limited at large angles. For the center beam, the smallest probe size achievable using these microscope parameters is roughly 40 $\textup{\AA}$ for an aperture with a diameter of $0.04 \, \textup{\AA}^{-1}$, corresponding to a semiangle of 0.4~mrads. This is smaller than a conventional uncorrected STEM experiment due to the large defocus applied (several \si{\micro\metre}) to reach the Scherzer condition.  Similarly, the outer ring aperture will form the smallest probe in real space for a radial size of $0.03 \, \textup{\AA}^{-1}$ and an annular size of $0.6 \, \textup{\AA}^{-1}$. However, as mentioned previously, the probe size is quite insensitive to the annular direction aperture size, which can be tuned from $0.1$ to $1 \, \textup{\AA}^{-1}$ without much change.

Fig.~\ref{Fig:Probes}d and e show the probe dimensions of the center and outer beams for two possible designs, where the intensity of all beams is approximately constant.  Fig.~\ref{Fig:Probes}d prioritizes the smallest possible outer beam, at the cost of making the center beam slightly larger.  Fig.~\ref{Fig:Probes}e makes the center beam smaller, at the cost of increasing the annular dimension of the outer beam probes. The latter configuration results in a more desirable set of operating conditions, where the probes are close to isotropic in the radial and annular directions, which will simplify the analysis. In practice however, this analysis is only used as a rough guideline, since the real apertures will never be fabricated perfectly, and the microscope will contain additional misaligned parameters (such as residual tilt, astigmatism, and coma) and aberrations not considered here. We recommend an iterative design process, where aperture geometries can be improved after testing with experiments. 

A similar design process can be used to design MBED apertures for aberration-corrected STEM instruments. The main difference is that in an aberration-corrected instrument, $C_3$ can be deliberately set to a negative value, which when combined with positive $C_1$ and $C_5$ aberration coefficients will lead to the flattest possible phase surface (and therefore smallest probes in real space) at large angles. This is similar to the ``negative Cs'' ($C_3$) concept for forming the smallest possible sub-$\rm{\AA}$ngstrom probe sizes \citep{jia2004high}.

% \subsection{MBED Experiment Design for an Aberration Correction}

% \hl{\bf Colin maybe add another theory section here}

%\subsection{Electron Diffraction of Crystalline Samples}

\section{Methods}

\subsection{Multislice Simulations}

All of the simulated diffraction experiments shown in this paper were generated using the methods and atomic potentials given by \cite{kirkland2020advanced}. The simulation algorithm employed was the multislice method \citep{cowley1957scattering}, as implemented in the Prismatic simulation code \citep{ophus2017fastSim, pryor2017streaming, rangeldacosta2020prismatic}; simulations were run in parallel through the use of GNU parallel \citep{tange_2020_3956817}. Simulations were performed at an acceleration voltage 200kV, with a beam convergence semiangle of 0.5 mrad. A single probe was placed at the center of the atomic coordinate cell. Atomic coordinates for the spherical Au nanoparticles were generated using pymatgen \citep{ONG2013314, togo2018}, and the atomic potential was sampled at a resolution of 0.125~$\textup{\AA}$ and a slice thickness of 2.0~\AA. The scattering potentials of individual atoms were radially integrated to 3.0~\AA. No thermal effects were applied to perturb the atomic coordinates. A total of 2387 particle orientations were simulated in order to sample diffraction behavior within the [001]-[101]-[111] symmetric projection of the cubic orientation sphere.

\subsection{FIB Amplitude Aperture Fabrication}

\begin{figure}[htbp]
    \centering
    \includegraphics[width=3.4in]{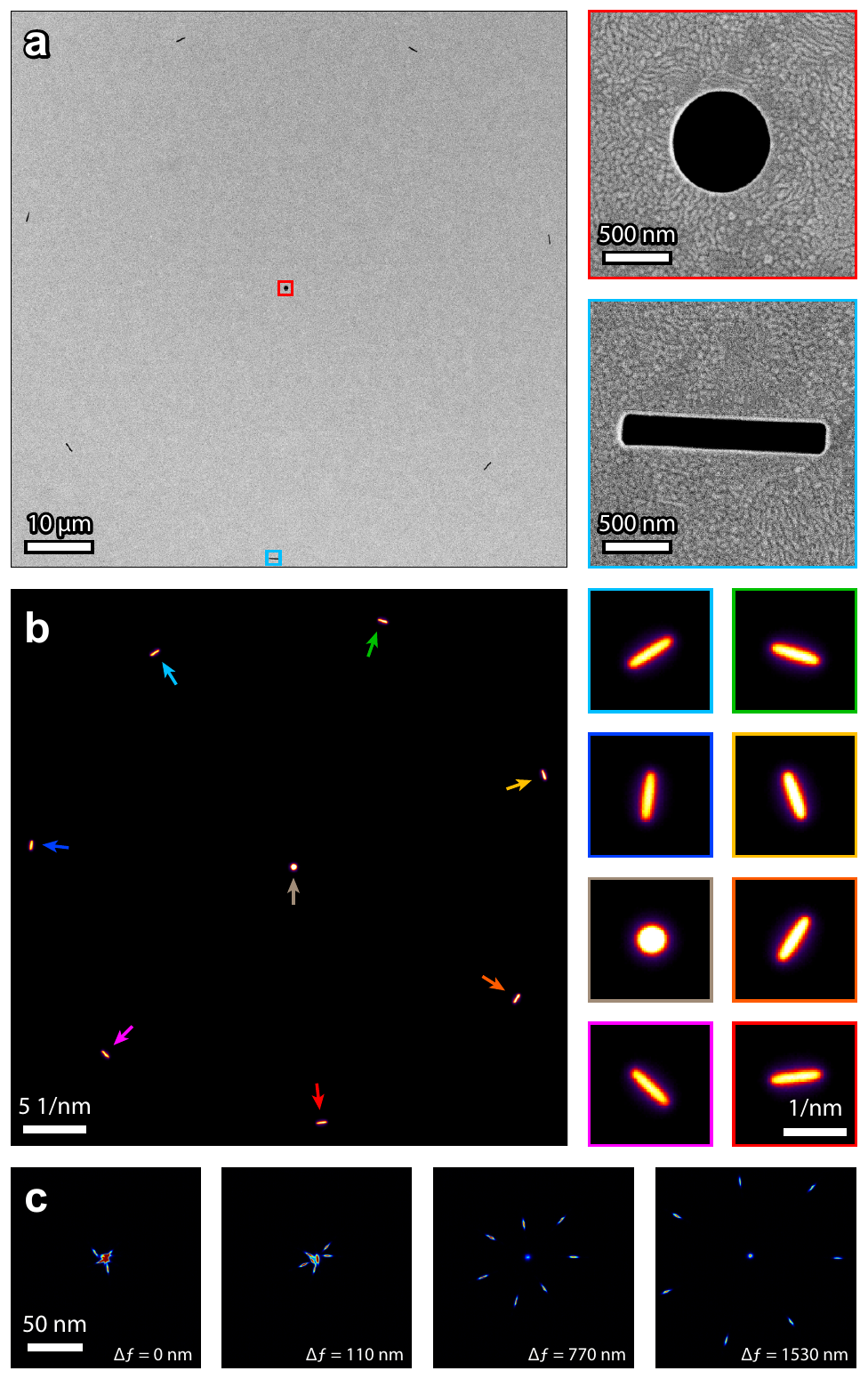}
    \caption{{\bf MBED aperture fabrication.} (a) SEM images of the apertures after FIB fabrication, with enlarged images of 2 of the apertures. (b) TEM diffraction images of the apertures, showing enlarged images of all apertures labeled with colored arrows. (c) TEM images in real space showing 8 beams generated by the apertures, at four different objective lens defocus values.}
    \label{Fig:apertures}
\end{figure}

The MBED apertures used in this study were fabricated in a silicon nitride membrane TEM window, purchased from Norcada, Canada. The single membrane on the window was $250 \, \si{\micro\metre} \times 250 \, \si{\micro\metre}$ square with a thickness of 200~nm. A 1 $\mu$m thick gold thin film was evaporated on one side of the whole window in order to make it opaque to electrons. The MBED aperture was milled into the membrane using a the focused ion beam column of an FEI Helios G4 UX system operating at 30kV with a current of 41~pA. Secondary electron images of the fabricated aperture are shown in Fig.~\ref{Fig:apertures}(a), and a diffraction image obtained with the aperture mounted in the TEM condenser system is shown in Fig.~\ref{Fig:apertures}(b). The aperture used for all experiments in this work has a ring radius of 80~\si{\micro\metre}, a center beam radius of 0.36~\si{\micro\metre}, and seven equally spaced outer beams with dimensions 1.52~$\times$~0.2~\si{\micro\metre}. %However, the smaller diameter aperture sets have the advantage that when forming probes with very small convergence angles, they can also limit the beam current when imaging beam-sensitive materials.

%that suit to scan the electron sensitive materials \textcolor{red}{(other advantages?)}.

\subsection{Experimental Setup and Alignment}

In order to test the MBED method for practical experiments, a set of apertures were installed on the C2 condenser aperture rod of a Thermo Fisher (FEI) TitanX TEM. This Titan was not equipped with an aberration corrector, and the spherical aberration $C_3$ is estimated to be 1.3 mm. Figure~\ref{Fig:apertures}b shows TEM diffraction space images of the aperture set, which are identical to the SEM images shown in Figure~\ref{Fig:apertures}a. Only the central round aperture and outer rectangular apertures allow electrons to pass though freely. The area outside the 8-aperture set was blocked by using an additional aperture in the C3 condenser position. The insets in Figure~\ref{Fig:apertures}b show the different rectangular shapes of the 7 outer beams, and the circular shape of the center beam. The different shapes of the diffraction space probes make it somewhat easier to distinguish the ``parent'' beam for diffraction experiments. However, in order to reduce the possible ambiguity of diffracted Bragg spots falling close to the mid-point between adjacent beams, we recommend using a large outer ring semi-convergence angle in order to reduce the overlap of adjacent patterns.

%The orientation of the 7 outer apertures are set differently for easily distinguishing diffraction patterns by the spots shapes from each other.

%\todo{The peak detection analysis does not bear this out...} And to keep the pattern separated far enough for the pattern divided purpose, the larger outer radius position apertures should be the first priority. 
%Meanwhile, large radius aperture has the great ability to generate large incident angle beams compared with the smaller ones, which will reduce the missing wedge for a better reconstruction processing. missing wadge. \textcolor{red}{7 fold apertures advantages.} Finally this 40um radius aperture set with the smallest aperture size is adopted into the next experiment step. \textcolor{red}{(I don't know if these explanation would duplicate in the previous section)}

%are the electron forbidden zone, which guarantees each nanobeam diffraction pattern is not affected by other electrons from the background. 

In order to align the MBED probe, we first set the lens currents in the condenser system of the microscope to select the desired $\alpha_{\textup{ring}}$ for the experiment. We then perform a standard STEM alignment using a circular aperture, and use the Ronchigram of the amorphous support film to bring the sample to eucentric height and correct the condenser astigmatism. Next we insert the MBED aperture plate and carefully center it on the optical axis. Slight misalignment of the aperture position is corrected by tilting the beam. We then fine-tune the condenser stigmators and the focus (using both the third condenser lens and the objective lens) to ensure that all of the outer beams are focused to the same point (astigmatism causes rays at different azimuths to be focused to a different plane). %We note that at large $\alpha_\text{ring}$  (greater than about 50~mrad on our instrument) the maximum current\todo{confirm this is right} allowed in the third condenser lens is not sufficient to focus the outer beams onto the optical axis at the eucentric plane, and so we must also tune the objective lens current to obtain the necessary focusing strength. 

For the experiments shown in this paper, we chose a semi-convergence angle of the outer ring of $\alpha_\text{ring} = 60$~mrad, giving a total angular coverage of the outer cone of beams of $6.8^\circ$. We used an accelerating voltage of 300 kV~for all experiments (i.e. $\lambda = 0.00197$ $\textup{nm}$), which positions the outer beams at a radius of $>38 \, \textup{nm}^{-1}$, outside the scattering vector limits for most samples. Figure~\ref{Fig:apertures}c shows images of the 8 probes in real space, at different defocus $\Delta f$ values. By tuning the defocus, we can produce an imaging condition with either the beams aligned to each other, or spatially separated. Note that even at $\Delta f = 0$ nm, the beams are not perfectly aligned due to residual stigmation and tilt in the condenser system. However, for MBED experiments where the probe step size is matched to the probe size (roughly 5~nm), this will not be an issue because the datasets generated from different beams can easily be aligned in post-processing. MBED datasets were acquired using a Gatan Orius CCD camera at 60~mm camera length with 1.5~pA probe current and 50~ms exposure time. 

%\subsection{Proof-of-Principle Samples}

\subsection{Sample Preparation}

The two experimental samples used in this study were both crystalline Au nanoparticles. The sample used for the MBED automated crystal orientation mapping (ACOM) experiments was produced by a chemical synthesis procedure \citep{nikoobakht2003preparation}. The sample used for the MBED tomography experiments was produced by evaporating Au onto a TEM grid with ultrathin carbon on a lacey support mesh.

\subsection{Bragg Peak Measurements}

% \hl{Ben you could add a Figure here if you wanted!  Up to you, I don't think we need it but would be a good excuse to show a BVM for MBED! -CO}

Most of the analysis performed in this paper used methods that are standard for NBED experiments \citep{ophus2014_4DSTEM, pekin2017strain, zeltmann2019patterned}. Measurement of the position of all diffracted Bragg disks was performed using the py4DSTEM analysis toolkit \citep{savitzky2020py4dstem}. First, we identified the positions of the 8 unscattered electron beams, and then constructed a template image of each of these 8 beams by cropping around them. Next, for each STEM probe position, we used cross correlation of the diffraction image with each of the 8 templates to identify candidate Bragg peak locations.

The next analysis steps are more specific to an MBED experiment.  In each image, we found the center position of all 8 unscattered beams. Each of the scattered beams was then assigned to a given parent beam, which was determined as the minimum distance to an unscattered beam. Note that this will produce a very small number of beams attributed to the wrong parent beam. In the case where the diffraction patterns from each beam overlap and make assignment of each Bragg disk to its parent beam ambiguous, they can be correctly assigned to a given parent beam by selecting the beam with the highest correlation intensity, since the correlation score will be maximum when the beams are the same shape. Here, we have taken a simpler approach and removed erroneous points by using a smaller radial cutoff around each parent beam. The final result is a list of all detected Bragg peaks, separated into different lists by each parent beam and STEM probe position.  The python notebook providing the above analysis, as well as the raw MBED data is available here  [link we be added after publication].

% [\hl{link we be added after publication - or we could just put it online now!}].

\subsection{MBED Orientation Mapping}

In order to determine the local orientation of the sample for a given MBED probe position, we matched a library of kinematic diffraction patterns to each MBED measurement, for all beams independently. This method is usually referred to as automated crystallographic orientation mapping (ACOM) in the literature \citep{rauch2014automated}. We computed a library of 8385 kinematic diffraction patterns of FCC Au following the methods given by \cite{degraef2003introduction, zuo2017advanced}. These patterns spanned an orientation range covering the symmetry-reduced triangular patch of the unit sphere from [001] to [011] to [111]. We assumed an accelerating voltage of 300 kV, and used a shape factor given by a 5 nm thick slab. We did not observe a very strong dependence of the results on the assumed thickness, and thus we used a relatively low thickness in order to include all possible diffraction spots for a given orientation. We have used a 1.3 $\textup{\AA}^{-1}$ cutoff in reciprocal space here.

To match each pattern, we developed a simple 2-step correlation scoring method. Initially, we compared the detected Bragg peak scattering vector lengths to the diffraction library lengths. We summed the absolute 1D distances between each simulated peak position and the nearest experimental peak. From this list, the simulated diffraction patterns with the lowest summed distances were selected for further analysis. The next step was to compare the subset of simulated patterns to the experimental patterns in 2D, rotating the simulated patterns 360 degrees in steps of 2 degrees. For each orientation, we computed a score given by the square root of the simulated peak intensity multiplied by the intensity of the experimental peak, within a cutoff threshold of 0.05 $\textup{\AA}^{-1}$. This sum over all simulated peaks was normalized by dividing by the total simulated peak intensity, excluding the center unscattered peak. This correlation score is effectively the normalized geometric mean of the simulated and experimental peak intensities, after pairing up each experimentally observed peak with a simulated peak where possible. This score reaches a maximum for the best fit pattern, which was then assigned as the local orientation. All orientation mapping was performed using custom Matlab scripts.

The above ACOM analysis was carried out for each beam individually, over all STEM probe positions. Once the orientation maps were computed for all 8 beams, the orientations of the outer beams were corrected by performing a 3D rotation inwards of 60 mrads to align the best-fit orientations of the outer beams to the center beam. In order to combine the orientation maps for multiple beams, we averaged the orientations by using a weighted median \citep{brownrigg1984weighted} of the 8 best-fit orientations, using correlation scores as the weights. In this way, beam orientations with low correlation scores were effectively removed from the final result.  The primary residual errors are due to overlapping crystallites in the sample, where either individual beams or the union of these beams contained diffraction signal from multiple orientations. These errors could perhaps be removed in the future by combining the orientation mapping with a 3D tomographic reconstruction, such at the method described in the next section.

\subsection{MBED 3D Tomographic Reconstruction}

Our MBED probes contain a center beam along the optical axis, and 7 beams arranged along a cone of angles tilted 60 mrad away from this optical axis. We therefore gain a small amount of 3D information about the sample, having covered an angular range of $\approx$6.8 degrees with the unscattered beams, and $\approx$10 degrees including the angular range of diffracted Bragg spots from most crystalline samples. Perhaps surprisingly, even these small angular ranges can provide useful information in TEM experiments. One example was shown by \cite{oveisi2017tilt}, who performed 3D reconstructions of dislocations using only 2 beams, separated by $1.56^\circ$. This experiment demonstrated that with sufficiently strong prior assumptions (in their case assuming a 1D curvilinear object), useful 3D information can be obtained from small angular ranges. In this paper, we demonstrate a simple proof-of-principle MBED tomography experiment by reconstructing just 2 scalar measurements, the virtual bright field and virtual dark field images. These were calculated by summing up the center beam correlation intensities to generate virtual bright field images, and all scattered Bragg peak correlation intensities to form virtual dark field images.

We employed the simultaneous iterative reconstruction technique (SIRT) algorithm to perform 3D tomgoraphic reconstructions \citep{gilbert1972iterative}. Because of the extremely large missing wedge present in MBED experiments, we regularize SIRT algorithm in two ways.  First, while we set the x-y dimensions of the reconstruction voxels to be equal to the STEM probe step size, we set the z dimension (along the optical axis) of the reconstruction voxels to be 10 times this size. The factor of 10 was chosen because the largest ray angles  from the optical axis for this aspect ratio is approximately $6^{\circ}$, close to the tilt range spanned by the experiment. This aspect ratio also reflects the much-reduced resolution available along the z direction due to the large missing wedge. Secondly, we apply a ``shrinkage'' regularization procedure, where the absolute value of the reconstructed voxels is reduced by a constant value after each SIRT update step \citep{parikh2014proximal, ren2020multiple}. We have also included an alignment step between each SIRT update, where the experimental images are aligned using cross correlation to the projected images from the reconstruction space. The resulting algorithm converges very rapidly to a sparse reconstruction, though it is somewhat sensitive to the strength of the regularization parameter. This reconstruction method was implemented using custom Matlab codes.

\begin{figure*}[htbp]
    \centering
    \includegraphics[width=6.2in]{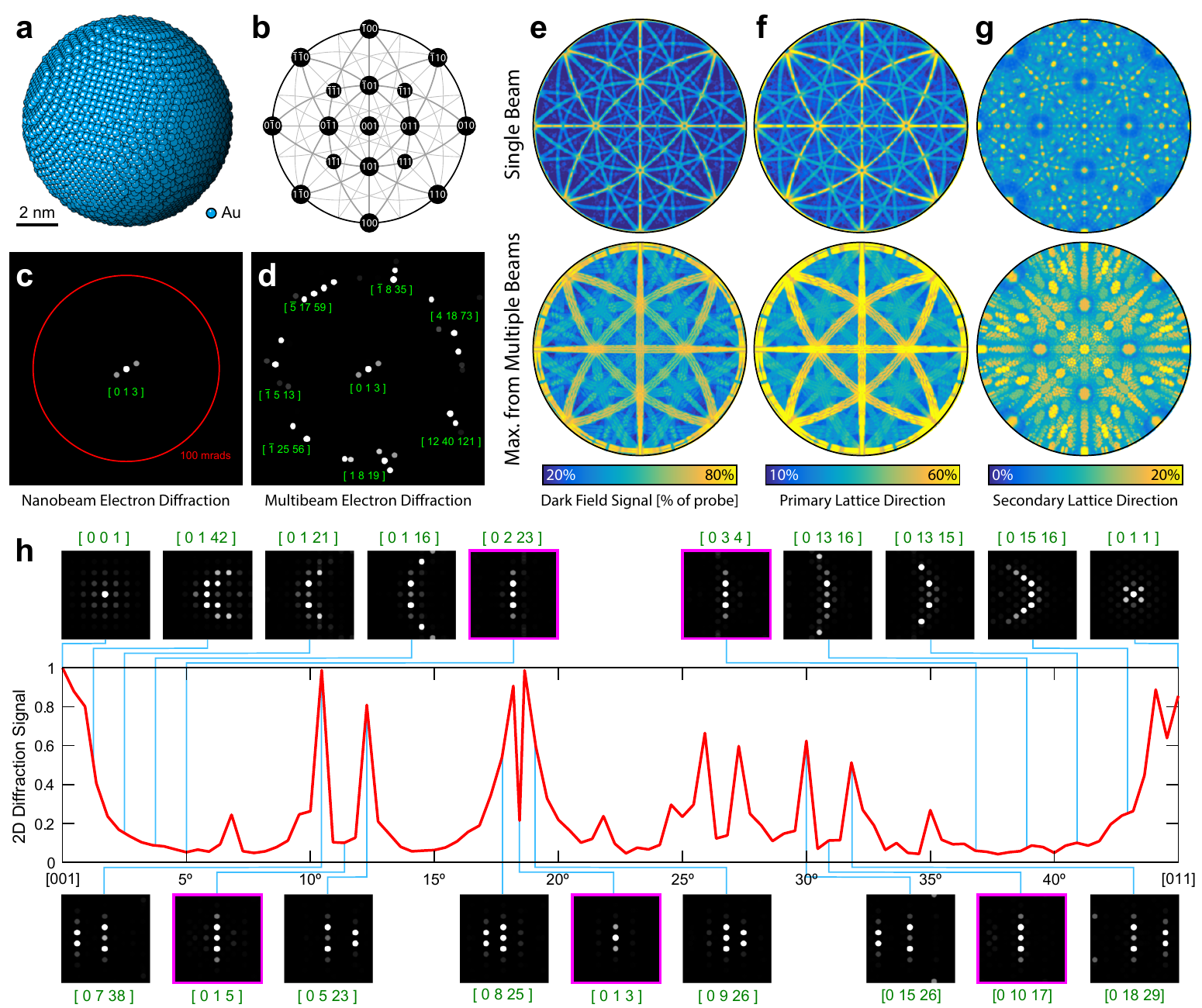}
    \caption{{\bf Multislice simulations demonstrating the benefits of MBED.} (a) Coordinates of a 10 nm diameter sphere of FCC Au atoms. (b) Legend of stereographic projections showing the nanoparticle orientation. (c) Single beam diffraction near the [013] zone axis, compared to (d) multibeam diffraction along the same zone axis with the outer beams along a 100 mrad ring, with the closest low index zone axis labels shown. (e) Total dark field signal for single beam and 8-beam MBED described in the text. Total 1D diffraction correlation signal along (f) the strongest scattering direction, and (g) perpendicular to the direction used in (f). (h) 2D diffraction signal as a function of particle orientation from [001] to [011] for the single beam case, with various selected diffraction patterns inset.  We have convolved all diffraction pattern images with a circular disk for easier visualization.}
    \label{Fig:sims}
\end{figure*}

\section{Results and Discussion}

\subsection{Simulations of Diffraction Orientation Dependence}

We performed dynamical multislice electron scattering simulations of MBED diffraction experiments for a single crystal at many different orientations in order to illustrate the potential benefits of MBED. Fig.~\ref{Fig:sims}a shows the sample geometry, a 10 nm diameter sphere of Au with the FCC crystal structure. Figs.~\ref{Fig:sims}c and d compare a conventional single-beam NBED experiment to an MBED experiment. In both simulations, the sample is oriented near to the [013] zone axis. The single beam experiment shows only 2 strongly excited Bragg peaks, with a 2 additional weakly excited peaks along the same direction. This pattern illustrates the primary issue with ACOM in TEM experiments: even when Bragg diffraction is present, many sample orientations lead to patterns that are essentially one-dimensional, and which cannot be used to uniquely identify the three-dimensional orientation vector [uvw] of the sample. By contrast, the MBED simulation shown in Fig.~\ref{Fig:sims}d contains significantly more orientation information. Each of the 8 patterns is labeled with the nearest low-index zone axis, and taken together give an unambiguous 3D orientation of [013].

We next quantify roughly how often we will encounter a case such as the diffraction pattern shown in Fig.~\ref{Fig:sims}c. Fig.~\ref{Fig:sims}b shows the orientation legend in a stereographic projection for the following figures. Fig.~\ref{Fig:sims}e compares the total intensity of the diffracted Bragg signals for the single-beam case to the maximum diffracted intensity across all 8 beams of an MBED experiment, for 1 center beam and 7 outer beams equally spaced around a 60 mrad ring. We note that the darker regions of the single beam case correspond to the orientations that do not produce strong diffraction signals; these regions are much reduced for the multibeam case. However, this analysis does not quantify how well we can overcome the ambiguity shown in Figs.~\ref{Fig:sims}c and d. 

In order to quantify whether a diffraction pattern can be uniquely indexed, we need to estimate how accurately diffraction vector lengths can be estimated in 2 orthogonal directions. We do this for each simulated diffraction pattern by first measuring the ``primary scattering direction'' of the diffraction pattern, given by the angle $\theta_0$ corresponding to the highest intensity $I(\mathbf{q})$ angular moment $M_\theta$
\begin{eqnarray}
    M_\theta &=& \sum I(\mathbf{q})
        \cos^2(\theta_q - \theta_0) \nonumber \\
    &=& 
        \sum I(\mathbf{q})
        \frac{[
        q_x \cos(\theta_0) + q_y \sin(\theta_0)
        ]^2}
        {|\mathbf{q}|^2},
\end{eqnarray}
where $\theta_q$ is the angular direction of the diffraction space coordinate system $\mathbf{q}$. Once we have determined this direction, we take the autocorrelation of the diffraction pattern
\begin{equation}
    I_{\textup{corr}}(\mathbf{q}) = 
    \mathcal{F}_{\mathbf{r}\to\mathbf{q}}
    \left\{
        | \mathcal{F}_{\mathbf{q}\to\mathbf{r}}
        \left\{
            I(\mathbf{q})
        \right\}|^2
    \right\},
\end{equation}
where $\mathcal{F}_{\mathbf{r}\to\mathbf{q}}$ and $\mathcal{F}_{\mathbf{q}\to\mathbf{r}}$ represent the forward and inverse 2D Fourier transforms. We then calculate the total correlation intensity along the primary and secondary directions equal to
\begin{eqnarray}
    I_{\textup{primary}} &=&
        \sum  I_{\textup{corr}}(\mathbf{q})
        \cos^2(\theta_q - \theta_0)
        \nonumber \\
    I_{\textup{secondary}} &=&
        \sum  I_{\textup{corr}}(\mathbf{q})
        \sin^2(\theta_q - \theta_0),
\end{eqnarray}
where $\theta_0$ is the direction which maximizes $M_\theta$ for each pattern. By definition, $I_{\textup{secondary}}$ cannot exceed $I_{\textup{primary}}$, and these 2 values should be equal for diffraction patterns which have 4-fold or mirror symmetry along both the $x$ and $y$ axis. In order to produce accurate results for patterns with a high degree of 6-fold symmetry, we have applied some low pass filtering to the $ I_{\textup{corr}}(\mathbf{q})$ images.

Fig.~\ref{Fig:sims}f shows $I_{\textup{primary}}$ as a function of orientation for both a single beam experiment, and the maximum of all 8 beams from an MBED experiment. These signals closely track the total dark field signals shown in Fig.~\ref{Fig:sims}(e). The diffraction signal along the secondary direction $I_{\textup{secondary}}$ is plotted as a function of orientation in Fig.~\ref{Fig:sims}(g), for both the single beam and MBED cases. These images have a very natural interpretation; a high value indicates strong diffraction signals in two dimensions, indicating diffraction patterns than should be easy to unambiguously index. All of the low-index zones $\langle uvw \rangle$ for FCC are represented by signal maxima, including $\langle 001 \rangle$, $\langle 011 \rangle$, $\langle 111 \rangle$, $\langle 012 \rangle$,  $\langle 112 \rangle$, $\langle 113 \rangle$, and others. The regions of low $I_{\textup{secondary}}$ signal indicate orientations that could produce ambiguous diffraction patterns, either with only diffraction vectors excited along one direction or no strong diffraction peaks. The total area fraction of these regions is much-reduced by MBED; the only large remaining orientations are those arranged in a cone around the $\langle 011 \rangle$ zone axes. These regions could be further filled in by increasing the angle of the outer beams beyond 60 mrads.

\begin{figure*}[htbp]
    \centering
    \includegraphics[width=6.2in]{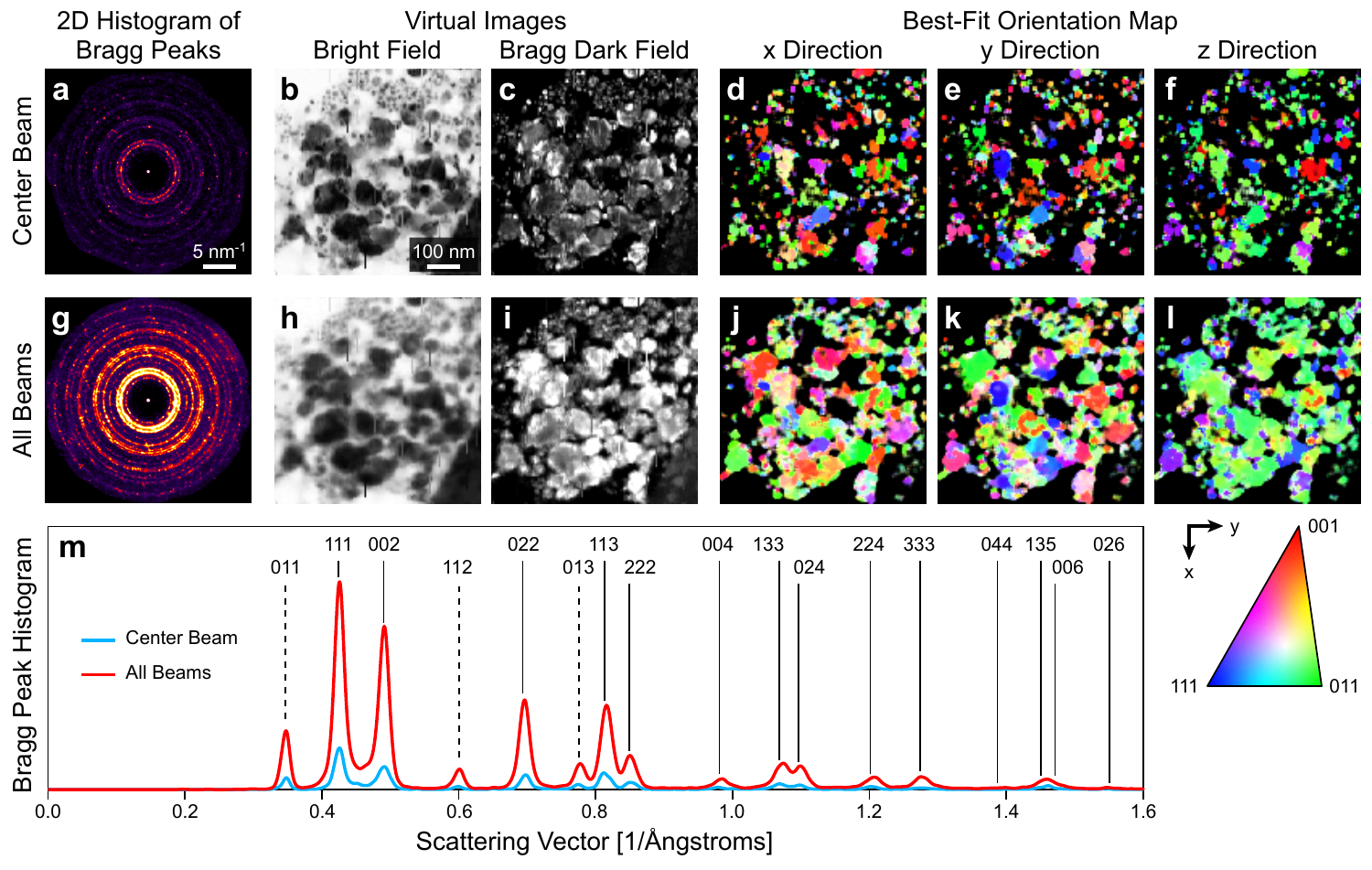}
    \caption{{\bf MBED orientation mapping of Au nanoparticles, comparing center beam to all beams.} (a,g) 2D histograms of all detected Bragg peaks. (b,h) Virtual bright field and (c,i) dark field images. (d-f) Orientation maps calculated from the center beam, and (j-l) orientation maps calculated from all beams. (m) Scattering vector histogram of all Bragg peaks, indexed to FCC Au.}
    \label{Fig:results_ACOM}
\end{figure*}

Finally, to show the potential orientation ambiguity in detail, we have plotted the ratio $I_{\textup{secondary}} / I_{\textup{primary}}$ in Fig.~\ref{Fig:sims}(h), following the orientations from [001] to [011]. As expected, the condition $I_{\textup{secondary}} = I_{\textup{primary}}$ is met or nearly met for high-symmetry zone axes such as the two endpoints.  Many other orientations also produce strong 2D diffraction signals, such as the insets shown for the [0 7 38], [0 5 23], [0 8 25], and [0 9 26] zone axes. However, the signal is extremely oscillatory, and many orientations produce a very low 2D diffraction signal.  In particular, we have highlighted 5 diffraction patterns with violet boxes, the [0 2 23], [0 1 5], [0 1 3], [0 10 17], and [0 3 4] orientations. These patterns are all essentially one-dimensional, and within the uncertainties given by varying particle thicknesses and shapes in a real experiment, they are effectively identical. Thus, these patterns cannot be assigned an unambiguous orientation from a single beam diffraction experiment. Fig.~\ref{Fig:sims}d however shows that MBED can be used to easily identify the true orientation in patterns that produce only 1D diffraction signals for single beam experiments.

\subsection{MBED Orientation Mapping of Polycrystalline Au}

In order to test the potential of MBED for crystal orientation mapping studies, we performed an MBED experiment on a sample consisting of highly-defected Au nanoparticles with sizes ranging from 5 - 100 nm. Fig.~\ref{Fig:results_ACOM}(a) shows a 2D histogram of the measured Bragg disk positions from the center beam of an MBED experiment covering a scan area of $128 \times 128$ STEM probe positions, with a step size of 5 nm. These Bragg peak intensities were used to construct virtual bright field and dark field images from the center beam and scattered beams respectively, which are shown in Figs.~\ref{Fig:results_ACOM}(b) and (c). These images show the wide size distribution and somewhat random morphology of the Au nanoparticles. Figs.~\ref{Fig:results_ACOM}(d-f) show the best-fit orientations in all 3 directions from the center beam. The orientation estimates have been masked by applying an intensity threshold to all diffracted beams, in order to prevent false positives from weak or non-existent diffraction patterns. Comparing these images to Fig.~\ref{Fig:results_ACOM}(b), we see many particles both large and small which have scattered enough to be visible in the bright field image, but could not be assigned a reliable orientation.

% not quite 8 times as many peaks due to the slightly weaker vacuum probe intensity of the outer beams\todo{what does weaker vac probe intensity mean?}. 
% More importantly than the total number of Bragg disks
% however, 

\begin{figure*}[htbp]
    \centering
    \includegraphics[width=6.2in]{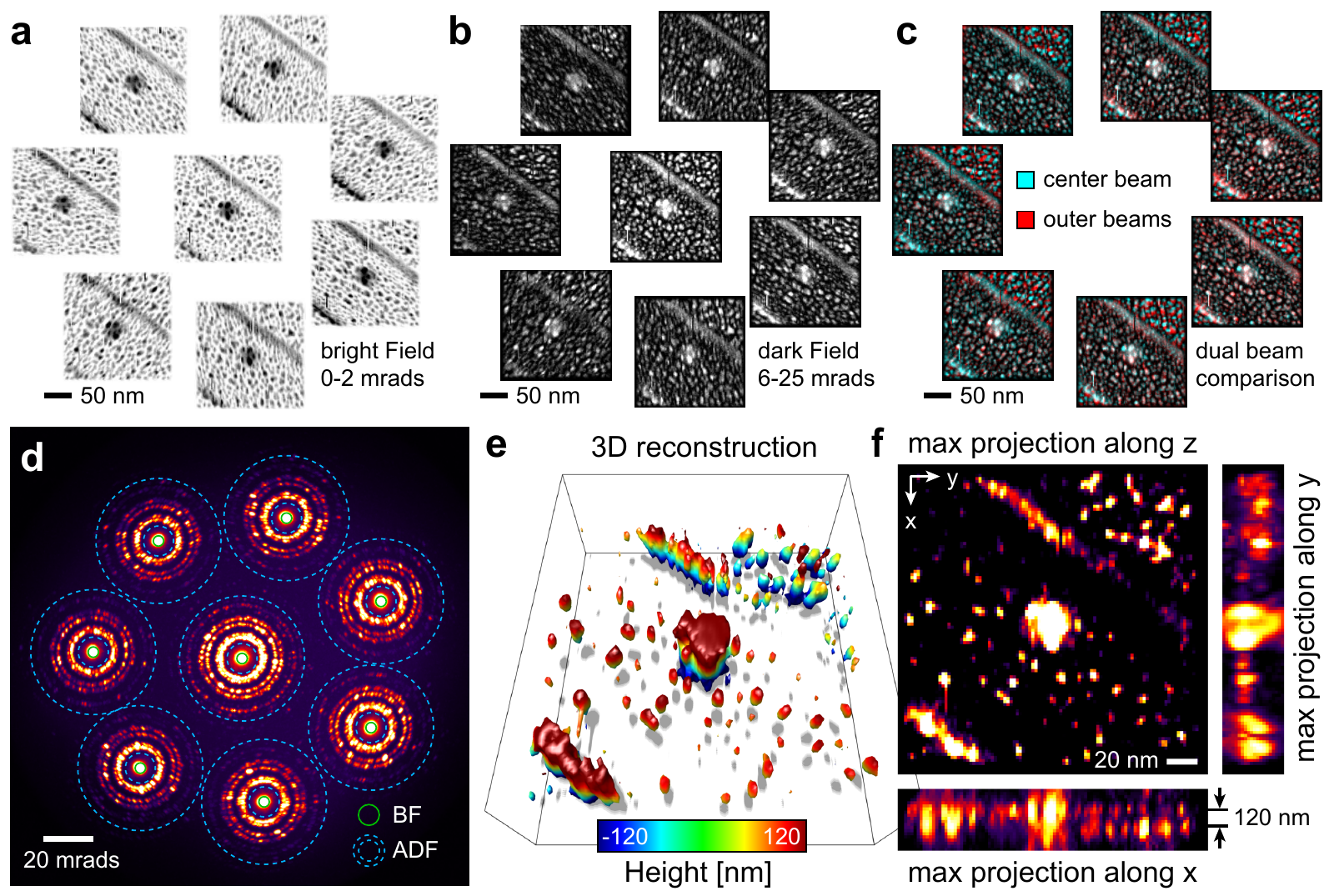}
    \caption{{\bf MBED 3D tomography of Au nanoparticles.} (a) Virtual bright field images for all beams. (b) Virtual annular dark field images from all beams. (c) Comparison between the center beam DF image and the DF images from each of the 7 outer beams. (d) Maximum CBED image from all probe positions, with virtual detector ranges overlaid. (e) 3D isosurface plot of the reconstruction. (f) Projections of the maximum intensity along the $x$, $y$, and $z$ directions of the reconstructed volume.}
    \label{Fig:results_tomography}
\end{figure*}

Fig.~\ref{Fig:results_ACOM}(g) shows a 2D histogram of the measured Bragg disk positions from all 8 beams of the MBED experiment. Approximately 5-6 times as many Bragg peaks were detected relative to the center beam measurements. This value is slightly smaller than the factor of 8 that would be expected for an 8-beam experiment. The reduction is due to the reduced intensity of the outer beams, due to their aperture areas being slightly reduced during the FIB machining process. However, the diffraction rings are much more contiguous in the entire MBED dataset; i.e.~many additional Bragg beams have been excited by the MBED measurement relative to a conventional NBED experiment. Figs.~\ref{Fig:results_ACOM}(h) and (i) show the virtual bright field and dark field images respectively, averaged over all MBED beams. The bright field image is virtually identical to the center beam shown in Fig.~\ref{Fig:results_ACOM}(b), though with some additional blurring to the slightly imperfect alignment of all beams. The virtual dark field image in Fig.~\ref{Fig:results_ACOM}(i) however shows significantly more constant signal across all grains; this is due to the much higher probability of Bragg excitation for all orientations in an MBED experiment. The orientation maps estimated from all beams (via weighted-median averaging of the 8 individual orientations) shown in Figs.~\ref{Fig:results_ACOM}(j-l) bear this out, with a significantly higher fraction of the pixels assigned to an orientation. The overall trend of the particle being biased towards an $\langle 011 \rangle$ orientation is also more clear.  The ACOM patterns are also now a very good match to the virtual images shown in Figs.~\ref{Fig:results_ACOM}(h-i); we can now estimate an orientation for virtually every pixel where the electron beam appears to be strongly scattered.

Fig.~\ref{Fig:results_ACOM}(m) shows a 1D histogram of the scattering vectors for all detected Bragg peaks, for both the center beam and all beams. All Bragg peaks that were observed could be indexed to FCC Au. All low index peaks occurred with probabilities relatively close to a random distribution, though with some preference for peaks such as (111) and (002) which can occur for a particle oriented along a [011] zone axis, the dominant orientation observed in Fig.~\ref{Fig:results_ACOM}(l). Three strongly excited forbidden reflections were also observed, indexed to (011), (112), and (013), and are marked by dashed lines in the figure. These peaks are likely due to the heavily defected nature of the crystalline grains in this sample. These reflections can also be used for the ACOM measurements, though they were not included in our diffraction library due to the use of kinematic structure factors (which are zero for these peaks).

\subsection{MBED 3D Reconstruction of Polycrystalline Au}

In order to test the potential of MBED experiments to recover 3D information from a sample, we acquired MBED data from $100 \times 100$ probe positions with a 2 nm step size from an alignment sample consisting of Au nanoparticles on ultrathin carbon with lacey support. Figs.~\ref{Fig:results_tomography}(a) and (b) show the virtual bright field and dark field images respectively.  The sample morphology consists of small Au particles on both the thin carbon membrane, and thicker carbon support (running diagonally across the field of view). This carbon support is moderately thick, and so should displace the Au nanoparticles out of the plane relative to the upper right and lower left corners. A cluster of Au nanoparticles near the center was used for the initial alignment. Unlike the previous section, these virtual images were constructed by using radial integration regions, shown for all beams in Figs.~\ref{Fig:results_tomography}(d). 

After completing the tomographic reconstruction, the beams were aligned more precisely to each other. This is shown in Figs.~\ref{Fig:results_tomography}(c) by directly overlaying each outer beam with the center beam, in 2 different color channels. In each of these images, the signals running across the carbon support are closely aligned. The upper right corner of the field of view shows strong parallax displacement however, where images of the nanoparticles in the outer beams show significant displacement relative to the center beam. Furthermore, these displacements are always in the radial direction for each beam, indicating that this is a true 3D effect, where the different beam angles intersect the sample at different positions for particles displaced out of the focal plane.

We performed a 3D tomographic reconstruction of the dark field images using the methods described previously. A 3D isosurface plot of the resulting  reconstruction is plotted in Fig.~\ref{Fig:results_tomography}(e). Each isosurface face is colored by the distance along the optical axis, over a range of 240~nm. The small Au nanoparticles show a strongly bimodal height distribution, where the particles on the support arm sit on a plane that is displaced from the plane occupied by the particles in the upper right and lower left of the field of view. This overall geometry is shown even more clearly in Fig.~\ref{Fig:results_tomography}f, especially the projection along the $x$ direction. Here, we can estimate the displacement of the 2 sets of particles to be 100-140~nm.

\section{Conclusion and Outlook}

% By using defocus to offset spherical aberration, we can focus these beams to the same point on the sample. 

In this paper, we have introduced multibeam electron diffraction, where we position a multiple probe-forming aperture below the second condenser lens of a STEM instrument. These apertures consist of a round center beam, and shaped outer beams falling on a ring of constant radius at a tilt angle which is large relative to the expected diffraction angles, similar to a precession electron diffraction experiment. We have demonstrated the utility of MBED for ACOM of polycrystalline samples using both multislice simulations and experimental measurements of Au nanoparticles. The large angular range of the beams in MBED excites significantly more Bragg diffraction and can thus pick up signals missed by conventional NBED experiments. We have also performed a proof-of-principle 3D tomographic reconstruction of Au nanoparticles using MBED. This experiment was able to show clear evidence of our ability to measure depth information, despite the angular range being small on an absolute scale. While precession electron diffraction can provide many of the same advantages for ACOM as MBED, it cannot provide 3D information due to the angular signals being mixed together. 

We believe MBED can improve virtually any NBED experiment that requires measurement of Bragg diffraction vectors, and so will be useful in many studies. The most promising next steps for development of the MBED technique are to combine crystallographic analysis such as ACOM with 3D tomographic reconstruction. It is possible to reconstruct a full 3D diffraction pattern in each voxel, as demonstrated by \cite{meng2016three}. While the resolution along the optical axis is much lower (both in real and reciprocal space) in MBED, our technique does not require multiple scans over the sample surface or scanning over multiple sample tilts. However, MBED could be easily combined with tilt series measurements, and will reduce the number of tilts that need to be collected due to its large range of angles collected in each scan.

Finally, we can also implement MBED in other ways. One idea would be to design MBED probes specifically for an aberration corrected instrument. One possible route would be to match the symmetry of the aperture layout to that of the non-radially-symmetric highest order aberrations. For a hexapole aberration corrector \citep{haider2008present}, this would mean placing 6 apertures on the outer ring, positioned along the directions with the flattest phase. With these design principles, aberration corrected MBED may be able to reach much larger angles than the 60 mrads demonstrated in this paper. To follow a design route targeting electron crystallography of beam-sensitive samples, we could design the beams to be offset from each other on the sample surface, in order to reduce the local dose to the sample. If these beams were offset primarily along a single direction, the measurements would be compatible with sample tilting such that each region of the sample would never see more than one beam. This could improve both micro- and nano-electron diffraction experiments that already use scanned electron probes \citep{gallagher2019nanoscale, bucker2020serial}. Any research group can easily pursue these ideas by following the simple fabrication route presented in this paper, using a FIB to machine custom apertures out of Si$_3$N$_4$ membranes with a thick metal layer deposited on one surface.

\section{Acknowledgements}

We thank Alex Bruefach and Mary Scott for providing the Au nanoparticle sample used for the MBED ACOM experiments. XH is support by National Key Research and Development Program of China (No. 2017YFA0303700), National Natural Science Foundation of China (No. 11504166), Molecular Foundry Proposal (No. 6463) and China Scholarship Council (No. 201906195020).  SEZ was supported by the National Science Foundation under STROBE Grant No. DMR 1548924. BHS and LRD are supported by the Toyota Research Institute. CO acknowledges support from the DOE Early Career Research Award program. Work at the Molecular Foundry was supported by the Office of Science, Office of Basic Energy Sciences, of the U.S. Department of Energy under Contract No. DE-AC02-05CH11231.

\section*{References}

\bibliographystyle{MandM}
\bibliography{4DSTEMrefs}

\end{document}